\documentclass[12pt]{article}
\usepackage{graphicx}

\begin{document}

\begin{center}

{\bf\LARGE Relativistic aberration of light as a\\
corollary of relativity of simultaneity\\}

\vskip 20pt

{\it{\textbf{Aleksandar Gjurchinovski
\footnote{E-mail: agjurcin@iunona.pmf.ukim.edu.mk, agjurcin@yahoo.com}}} \\
\vskip 5pt
Department of Physics, Faculty of Natural Sciences \\ and Mathematics,
Sts. Cyril and Methodius University \\ P. O. Box 162, 1000 Skopje, Macedonia}
\end{center}

\vskip 20pt

\begin{center}
{\bf\large Abstract}
\end{center}

\begin{quote}
A new derivation of the relativistic aberration formula for a
plane-polarized light wave is presented that does not require
any use of the Lorentz transformation. 
The method is based on a modification of the Huygens-Fresnel
principle to include the relativistic effects introduced by
the relative motion between the observer and the emitter.
The derivation clearly shows that the aberration formula is a 
direct consequence of the relative simultaneity.
 
\vskip 5pt
{\bf PACS Numbers:} 42.15.Dp, 42.15.Fr, 03.30.+p
\end{quote}

\section{Introduction}

Relativistic aberration of light is a change of the direction
of a light ray when observed from various inertial reference 
frames \cite{liebscher,muller,jackson,landau,gjurchinovski}.
It is usually, and sometimes incorrectly, discussed in the
context of stellar aberrations \cite{phipps,eisner,kassner}.
The traditional derivation of the relativistic aberration 
formula involves a plane-polarized electromagnetic wave
and a requirement that its phase is an invariant quantity 
under a Lorentz transformation \cite{muller,jackson}. 
The aberration formula also can be obtained by a direct 
application of the addition law for relativistic
velocities \cite{landau}.

The purpose of the present paper is to provide a novel insight
into the problem of relativistic aberration of light.
We will investigate the case of a plane-polarized light-wave
emanating from a uniformly moving emitter. 
As a simple model of a plane-wave emitter we use a two-dimensional
array of identical coherent elementary sources equally and 
densely distributed on a given plane $\Sigma$. 
We take the elementary sources to start radiating spherical 
wavelets simultaneously with respect to the reference frame 
where the plane $\Sigma$ is at rest. 
By letting the distance between each two adjacent elementary 
sources to approach zero, we may consider this 
two-dimensional arrangement as an ideal plane-wave emitter.

In Section 2 we derive the formula for the aberration of a vertical 
light ray as a simple case of a relativistic aberration 
usually discussed in the context of the Michelson-Morley 
experiment \cite{schumacher,gjurchinovski2}.
As a tool for explaining the formation of the aberrated wavefronts, 
we use a relativistic modification of the Huygens-Fresnel principle 
on the spherical wavelets radiated by the elementary sources from
the emitter.
We take into account that the elementary sources will not flash 
simultaneously in the reference frame where the emitter is moving
by making a simple analogy with the famous Einstein's
train-embankment thought experiment \cite{janis}.
The described procedure is repeated in Section 3 for a more general 
situation when the light ray emanates at an arbitrary angle 
from the emitter.

\section{Aberration of a vertical light ray}

Figure 1 depicts a special case of a horizontal plane-wave 
emitter. 
A simple Huygens' construction shows that this idealized 
emitter will radiate plane wavefronts at a vertical direction.
However, the situation is changed with respect to the frame
in which the emitter is moving at a constant speed $v$ to 
the right (see Fig. 2).
Due to the motion of the emitter, the elementary sources along 
the emitter will not flash simultaneously, but continously 
from left to right.
As a consequence of this simultaneity-loss, there will exist 
a certain non-zero time interval between the beginning of the 
emission of any two elementary sources on the emitter,
and this time interval will depend on the distance 
between the sources.

Let $A$ and $B$ denote two arbitrary chosen elementary sources
that belong to the emitter $\Sigma$, and let $l=\overline{AB}$ be
the distance between the sources.
If the elementary sources $A$ and $B$ started to emit 
spherical wavelets simultaneously in the reference frame where
the emitter is at rest, then in the reference frame
where the emitter is moving at a constant speed $v$ to the right
(Fig. 2), the elementary source at $B$ will start radiating
wavelets $\Delta t_{AB}$ seconds after the flash of the elementary
source at $A$. The time interval $\Delta t_{AB}$ between the 
emission of the elementary sources is given by
\begin{equation}
\Delta t_{AB}= {lv \over c^2-v^2},
\label{eq:1}
\end{equation}
where $c$ is the speed of light in vacuum. 
Equation (\ref{eq:1}) follows directly from the postulates of
special relativity if one makes an analogy with Einstein's 
train-embankment thought experiment in which the opposite ends 
of the train are hit by strokes of lightning by simply substituting 
the train with the plane-wave emitter, and the strokes with the 
flashes of the elementary sources $A$ and $B$. In order to better 
establish this analogy, the reader is referred to the derivation
of Eq. (2.3) in the paper by Janis \cite{janis}.

During the time-interval $\Delta t_{AB}$ taken from the beginning
of the radiation of the source at $A$, the source at $B$ has moved 
the distance $\overline{BB'}=v\Delta t_{AB}$ to the right. 
At the end of this time-interval $\Delta t_{AB}$, the source at $B$
(now at the point $B'$) will begin to radiate, and the elementary 
wavefront radiated from $A$ a time $\Delta t_{AB}$ earlier will be
a sphere of radius $\overline{AC}=c\Delta t_{AB}$.
We have taken into account the constant-light-speed postulate
according to which the elementary wavefront radiated by a moving
elementary source is a sphere expanding in all directions at a 
constant speed $c$.

The wavefront $\Omega$ emitted by the plane-wave emitter
$\Sigma$ at a given instant of time is constructed as an 
envelope of all the spherical wavelets from the point sources 
on the emitter that have radiated until that instant. 
At the time when the source at $B'$ flashes, the wavefront $\Omega$ 
is an envelope of the spherical wavelets from the point sources 
that are located at left from the point $B'$, since the point 
sources at right from $B'$ have not flashed yet. 
However, at later times, the sources at right from the point $B'$ 
will also participate into the formation 
of the propagating wavefront.

From Fig. 2 we see that the wavefront $\Omega$ will 
propagate at a speed $c$ at an angle $\theta$ from the vertical.
From the triangle $AB'C$, the angle $\theta$ can be expressed as
\begin{equation}
\sin\theta={\overline{AC}\over\overline{AB}+\overline{BB'}}=
{c\Delta t_{AB} \over l+v\Delta t_{AB}}.
\label{eq:2}
\end{equation}
We substitute Eq. (\ref{eq:1}) into Eq. (\ref{eq:2}) and obtain
\begin{equation}
\sin\theta={v\over c},
\label{eq:3}
\end{equation}
which is the well-known formula that describes the aberration of 
the transverse light ray in the Michelson-Morley 
apparatus \cite{schumacher,gjurchinovski2}.

\section{A derivation of the general formula for relativistic
aberration of light}

In the following we will consider the general situation of 
relativistic aberration when the plane-wave emitter is 
inclined at an arbitrary angle $\theta_0$ to the horizontal with 
respect to its rest frame of reference (Fig. 3).
Evidently, the plane-wave emitter will emit plane wavefronts at 
an angle $\theta_0$ from the vertical. 
In the reference frame where the emitter is moving at a constant
speed $v$ to the right (see Fig. 4), its inclination angle 
$\psi$ will differ from its inclination angle $\theta_0$ in 
the stationary case due to the effect of Lorentz contraction
along the direction of its motion. This tilt of the moving 
emitter is given by
\begin{equation}
\tan\psi={\tan\theta_0\over \sqrt{1-v^2/c^2}}.
\label{eq:4}
\end{equation}
Furthermore, the elementary sources along the plane-wave emitter
which simultaneously started their radiation in the emitter's
rest reference frame will not flash simultaneously in the
reference frame where the emitter is moving.
If we choose two elementary sources $A$ and $B$ on the emitter and
follow the same considerations as in Section 2, we may
argue that the time-interval $\Delta t_{AB}$ that
elapsed between the flashes of the elementary sources $A$ and
$B$ from the moving emitter can be calculated as
\begin{equation}
\Delta t_{AB}={lv \over c^2-v^2}\cos\psi,
\label{eq:5}
\end{equation}
where $l=\overline{AB}$ is the distance between the sources.
We take into account that the elementary source $B$ will 
start radiating spherical wavelets $\Delta t_{AB}$ seconds 
later than the elementary source $A$. Due to the motion
of the emitter, the flash of the source $B$ will occur 
from a different position in space, that is, from the 
space-point $B'$.
From Fig. 4 we see that during this time-interval 
$\Delta t_{AB}$, the emitter has covered the distance 
$\overline{AA'}=\overline{BB'}= v\Delta t_{AB}$ 
to the right, and the wavelet emitted from $A$ has evolved 
into a spherical wavefront of radius $\overline{AC}=c\Delta t_{AB}$.
The plane wavefront $\Omega$ emitted by the moving 
plane-wave emitter at the end of this time-interval is an 
envelope of all the spherical wavelets that have been radiated until 
that instant, and it propagates at a speed of light $c$ at an 
angle $\theta$ from the vertical.
From the triangle $ACD$ we have
\begin{equation}
\sin\theta={\overline{AC}\over \overline{AA'}+\overline{A'D}}=
{c\Delta t_{AB}\over v\Delta t_{AB}+\overline{A'D}}.
\label{eq:6}
\end{equation}
Applying the sine theorem to the triangle $A'B'D$, we get
\begin{equation}
\overline{A'D}=l{\sin(\theta-\psi)\over \sin\theta}.
\label{eq:7}
\end{equation}
We substitute Eqs. (\ref{eq:5}) and (\ref{eq:7}) into Eq. (\ref{eq:6})
and simplify the result to obtain
\begin{equation}
\left( 1-{v^2\over c^2} \right) \cos\theta\tan\psi=
\sin\theta-{v\over c}.
\label{eq:8}
\end{equation}
By taking the square of Eq. (\ref{eq:8}) and rearranging the
terms, we obtain a quadratic equation in $\sin\theta$:
\begin{equation}
\left[ 1+\left(1-{v^2\over c^2}\right)\tan^2\theta_0\right]
\sin^2\theta-2{v\over c}\sin\theta+{v^2\over c^2}-
\left( 1-{v^2\over c^2}\right)\tan^2\theta_0=0,
\label{eq:9}
\end{equation}
where we also have taken into account Eq. (\ref{eq:4}). 
Equation (\ref{eq:9}) has two solutions in $\sin\theta$:
\begin{equation}
(\sin\theta)_{1,2}={(v/c)\cos^2\theta_0\pm(1-v^2/c^2)
\sin\theta_0\over 1-(v^2/c^2)\sin^2\theta_0}.
\label{eq:10}
\end{equation}
From the requirement $\theta=\pi/2$ when $\theta_0=\pi/2$ (i.e. 
the direction of the light ray will remain unchanged with 
respect to any inertial observer that moves at a
constant velocity in the direction of the ray),
we find the only solution of Eq. (\ref{eq:9}) that correctly
describes the propagation of the wavefront $\Omega$:
\begin{equation}
\sin\theta={\sin\theta_0+v/c\over 1+(v/c)\sin\theta_0}.
\label{eq:11}
\end{equation}
Equation (\ref{eq:11}) is the law of aberration in its
most general form, and it is identical to the formula obtained
with the standard methods \cite{muller,jackson,landau}.
Obviously, in the case $\theta_0=0$,
Eq. (\ref{eq:11}) reduces to Eq. (\ref{eq:3}).

\section{Concluding remarks}

We have presented a novel approach to the problem of relativistic
aberration of light by using a modification of the Huygens-Fresnel
principle to incorporate the relativistic effects due to the 
relative motion between the source and the observer.
The resulting formulas are in perfect agreement with the ones obtained 
with the standard techniques.
The method of derivation is suitable for introductory physics courses,
and it can serve as an explicit demonstration of
the fact that the relativistic aberration of light is a direct 
consequence of the relativity of simultaneity.
The derivation also can be considered as a proof of the validity 
of the Huygens-Fresnel principle in the case of a rapidly moving 
wave-emitter if the relativistic effects are taken into account.

As a subject for future study, we leave to the reader to explore the 
situations when the elementary sources are arranged in a more
complicated manner. 
For example, an interesting case occurs when the elementary sources are
arranged on a spherical surface, which in the limit of infinitesimally 
small distances between the sources represents an ideal 
three-dimensional model of a spherical-wave emitter.


\newpage

\begin{center}
   \includegraphics[width=5in]{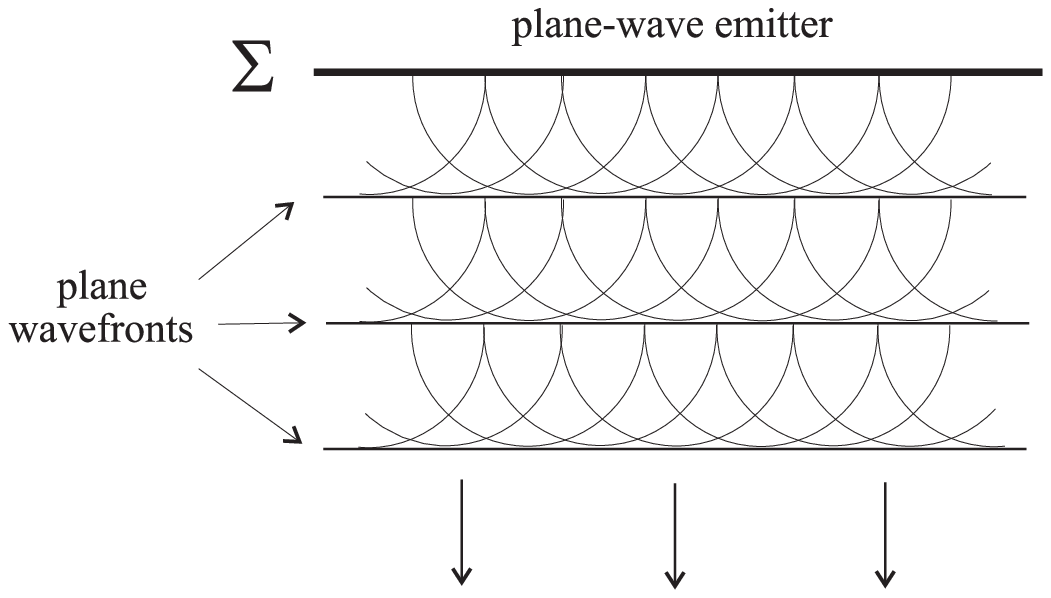}
\end{center}

\noindent{\bf Figure 1.}
The wavefronts radiated by the plane-wave emitter $\Sigma$ will
propagate from the emitter at a vertical direction. 
Here we consider only the wavefronts that are radiated vertically
downward.

\newpage

\begin{center}
   \includegraphics[width=5in]{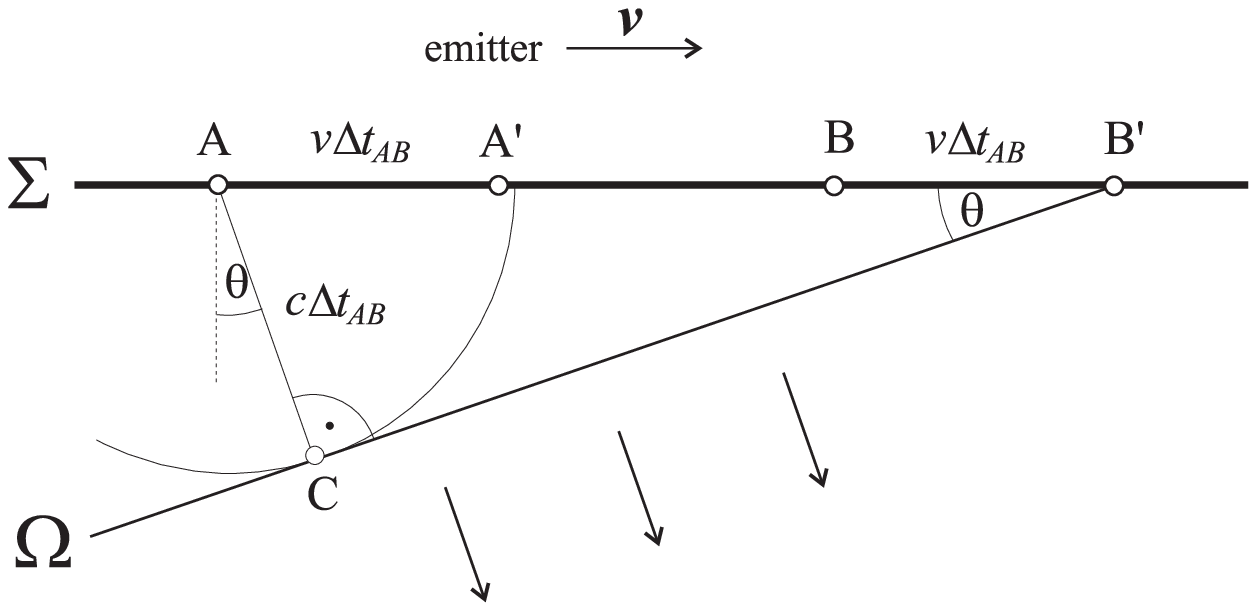}
\end{center}

\noindent{\bf Figure 2.}
Huygens' construction of the wavefront $\Omega$ with respect 
to the frame where the emitter $\Sigma$ is moving at a constant 
speed $v$ to the right. 
The plane wavefront $\Omega$ will propagate at a speed of light 
$c$ at an angle $\theta$ from the vertical.
Observe that the elementary source $B$ will start radiating 
spherical wavelets at the space-point $B'$ a time $\Delta t_{AB}$
later than the elementary source $A$.

\newpage

\begin{center}
   \includegraphics[width=5in]{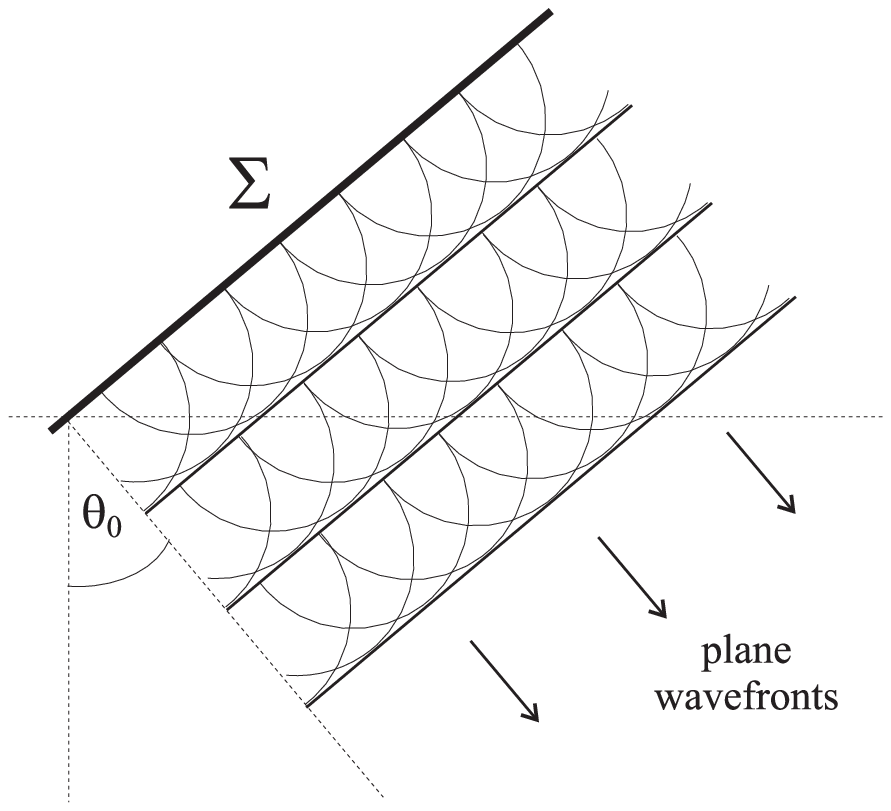}
\end{center}

\noindent{\bf Figure 3.}
Several consecutive wavefronts emitted by a stationary plane-wave
emitter $\Sigma$ inclined at an angle $\theta_0$ to the horizontal. 
The plane wavefronts will propagate at an angle $\theta_0$ from
the vertical. 

\newpage

\begin{center}
   \includegraphics[width=5in]{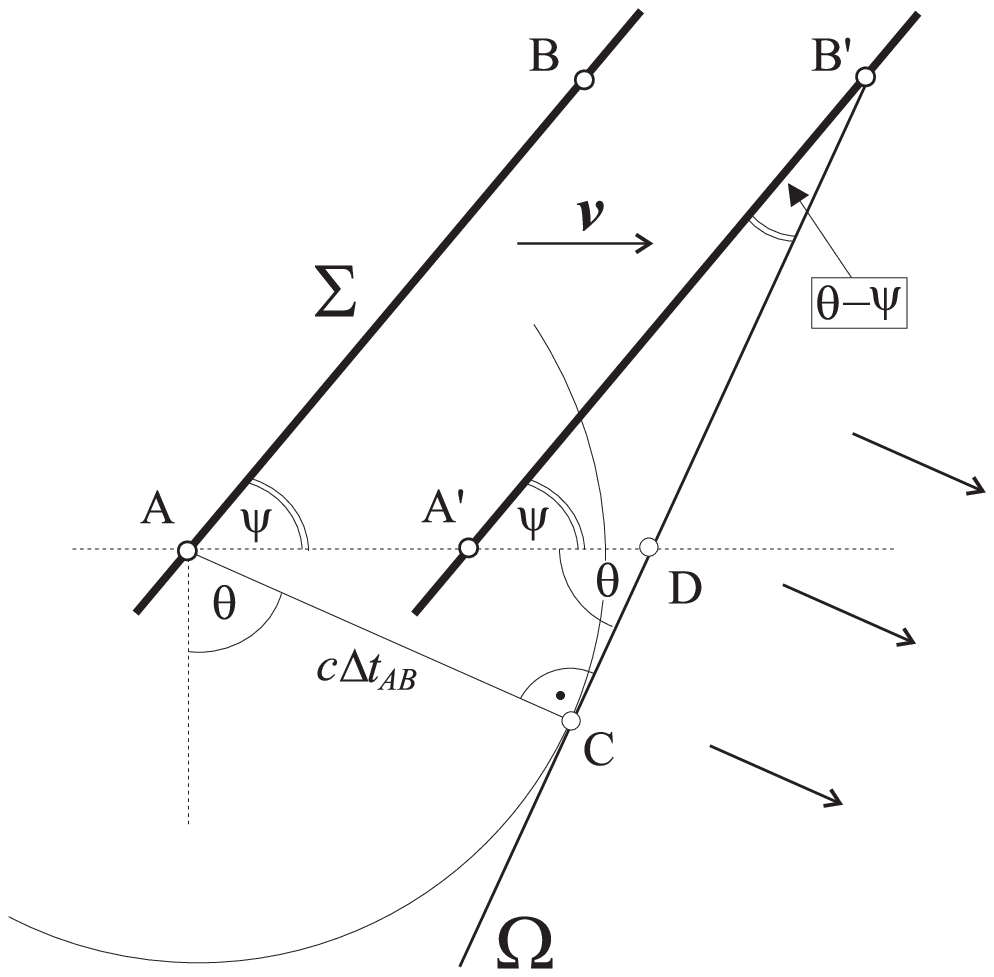}
\end{center}

\noindent{\bf Figure 4.}
In the frame where the emitter $\Sigma$ is in uniform rectilinear
motion at a speed $v$ to the right, its inclination angle $\psi$
will differ from its inclination angle $\theta_0$ in the stationary
case due to the Lorentz contraction along the direction of its motion.
The plane wavefront radiated by the moving emitter is a tangent 
of the spherical wavelets, and it propagates
at a speed of light $c$ at an angle $\theta$ from the vertical.


\begin{thebibliography}{11}

\bibitem{liebscher}
D. E. Liebscher and P. Brosche, "Aberration and relativity,"
Astron. Nachr. {\bf 319}, 309-318 (1998).

\bibitem{muller}
C. M\"oller, {\it The Theory of Relativity} (Clarendon Press,
Oxford, 1972), 2nd ed.

\bibitem{jackson}
J. D. Jackson, {\it Classical Electrodynamics} (Wiley, New York,
1975), 2nd ed.

\bibitem{landau}
L. Landau and E. Lifshitz, {\it The Classical Theory of Fields}
(Addison-Wesley, Cambridge, 1951).

\bibitem{gjurchinovski}
A. Gjurchinovski, "Aberration of light in a uniformly moving
optical medium," Am. J. Phys {\bf 72}, 934-940 (2004).

\bibitem{phipps}
T. E. Phipps, Jr., "Relativity and aberration," Am. J. Phys. 
{\bf 57}, 549-551 (1989).

\bibitem{eisner}
E. Eisner, "Aberration of light from binary stars -- a paradox?,"
Am. J. Phys. {\bf 35}, 817-819 (1967).

\bibitem{kassner}
K. Kassner, "Why the Bradley aberration cannot be used to measure
absolute speeds. A comment.," Europhys. Lett. {\bf 58}, 
637-638 (2002).

\bibitem{schumacher}
R. A. Schumacher, "Special relativity and the Michelson-Morley
interferometer," Am. J. Phys. {\bf 62}, 609-612 (1994).

\bibitem{gjurchinovski2}
A. Gjurchinovski, "Reflection of light from a uniformly moving
mirror," Am. J. Phys. {\bf 72}, 1316-1324 (2004).

\bibitem{janis}
A. I. Janis, "Simultaneity and special-relativistic kinematics,"
Am. J. Phys. {\bf 51}, 209-213 (1983).

\end{thebibliography}
\end{document}